\documentstyle[epsfig]{article}

\newcommand{\lnst}{\hspace*{-\parindent} }
\newcommand{\ihsp}{\hspace*{\fill} }

\newcommand{\case}[2]{\mbox{\small $\displaystyle \frac{#1}{#2}$}}

\newcommand{\etal}{{\it et al.}}
\newcommand{\bcn}{\begin{center}}
\newcommand{\beq}{\begin{equation}}
\newcommand{\beqar}{\begin{eqnarray}}

\newcommand{\Pslash}{\mbox{$\not \! P$}}

\newcommand{\qslash}{\mbox{$\not \! q$}}
\newcommand{\kslash}{\mbox{$\not \! k$}}

\newcommand{\ecn}{\end{center}}
\newcommand{\eeq}{\end{equation}}
\newcommand{\eeqar}{\end{eqnarray}}

\newcommand{\Eq}[1]{Eq.~(\ref{#1})}

\newcommand{\Fig}[1]{Fig.~\ref{#1}}

\newcommand{\Table}[1]{Table~\ref{#1}}

\begin{document}

\parbox{4.9in}{ 
\footnotesize
Contribution to the International Conference on Nuclear and
Particle Physics with CEBAF at Jefferson Lab, November 1998, Dubrovnik,
Croatia.  Proceedings to be published. \hspace*{\fill}
\parbox[t]{25mm}{KSUCNR-101-99 \\ }  }
\normalsize

\bcn
{\bf ELECTROMAGNETIC FORM FACTORS OF MESON TRANSITIONS}
\medskip

PETER C. TANDY\\
{\em Center for Nuclear Research, Department of Physics\\
Kent State University, Kent, OH, USA 44242} 
\ecn

\lnst Within the Dyson-Schwinger equation approach to 
modeling QCD for meson physics, we present new results for 
$g_{\rho\pi\pi}$ and the coupling constants and form factors for 
the transitions  $\gamma^\ast \pi \rho$ and 
$\gamma^\ast \pi^0 \gamma$.    We discuss the role of the sub-dominant 
covariants of the $\pi$ Bethe-Salpeter amplitude and investigate the 
asymptotic behavior of the $\gamma^\ast \pi^0 \gamma$ form 
factor. 

\smallskip
\lnst {\it PACS numbers}: 12.38.Lg, 13.40.Gp, 14.40.Aq,
24.85.+p \vspace*{0.1em}

\lnst {\it Keywords}: Nonperturbative modeling of QCD, Dyson-Schwinger 
Equations, Confinement, Meson coupling, Rho and pion electromagnetic 
transitions, Form factors

\medskip
\bcn
{\large\it 1.~QCD Modeling of Mesons}
\ecn

The  Dyson-Schwinger equation (DSE) approach~[\ref{DSErev}] to 
non-perturbative QCD modeling of hadrons and their interactions~[\ref{T97}]
combines  truncated QCD equations of motion for propagators and vertices 
with infrared phenomenology fitted to a few key low-energy observables.
Parameter-free predictions for other hadron properties and observables 
are then used to develop and test our understanding of hadron physics
at the quark-gluon level.  The DSE for the fully-dressed and 
renormalized quark propagator in Euclidean metric is
\begin{equation}
 S^{-1}(p) = Z_2 [i \gamma\cdot p + m_0(\Lambda)]
     + Z_1 \; \case{4}{3} \int^\Lambda \frac{d^4k}{(2\pi)^4} g^2 
D_{\mu \nu}(p-k)
        \gamma_\mu S(k) \Gamma_\nu^g (k,p) ,     \label{fullDSE}
\end{equation}
where $m_0(\Lambda)$ is the bare mass parameter and $\Lambda$ 
characterizes the regularization mass scale. In the DSE approach the 
renormalized dressed gluon propagator
$D_{\mu\nu}(q)$  and dressed quark-gluon vertex $\Gamma_\mu^g(k,p)$ 
are constrained in the UV by perturbative results and are representated
by phenomenological IR forms with parameters fitted to selected pion 
and kaon observables.

Mesons are generated as bound states of a quark of flavor $f_1$ 
and an antiquark of flavor $\bar f_2$ via the 
Bethe-Salpeter (BS) equation
\begin{equation}
\Gamma(p;P) =  \int \frac{d^4q}{(2\pi)^4} K(p,q;P) S_{f_1}(q + \xi P) 
                     \Gamma(q;P) S_{f_2}(q - \bar\xi P)   ~, \label{bspi}
\end{equation}
where \mbox{$\xi + \bar\xi = 1$} describes momentum sharing. 
The kernel $K$ is the renormalized, amputated $\bar q q$ scattering
kernel that is irreducible with respect to a pair of $\bar q q$ lines. 
The present stage of QCD modeling truncates $K(p,q;P)$ at the ladder 
approximation and couples quark color currents with bare vertices and 
an effective gluon 2-point function in Landau gauge.  The latter 
involves  $\alpha_{\rm eff}(q^2)$ for interpolation between the 1-loop
pQCD result in the UV and a phenomenological enhancement in the IR.
The treatment of the quark DSE that is dynamically matched to this is
the bare vertex or rainbow approximation; the axial vector 
Ward-Takahashi identity is then preserved, Goldstone's theorem is 
manifest, and realistic pion and kaon solutions are ensured~[\ref{MR97}].

The general form of the pion BS amplitude is
\beq
\hat{\Gamma}_\pi^j(k;P)  =   \tau^j \gamma_5 [ i E_\pi(k;P) + 
\Pslash F_\pi(k;P) 
+ \kslash \,k \cdot P\, G_\pi(k;P) 
+ \sigma_{\mu\nu}\,k_\mu P_\nu \,H_\pi(k;P) ]\, ,
\label{Gammapi}
\eeq
and the first three terms are significant in realistic 
solutions~[\ref{MR97}].   With \mbox{$S^{-1}(p)=$}
\mbox{$i\kslash A(k^2) + B(k^2)+ m$}, the axial Ward-Takahashi 
identity in the chiral limit specifies the dominant amplitude as
\mbox{$E_\pi(k;P=0)=$} \mbox{$B_0(k^2)/f_\pi$}; the amplitudes 
$F_\pi$ and $G_\pi$ are related to $A(k^2)$ 
and its derivative in a less direct manner~[\ref{MRT98}].  

In Euclidean metric, the BS meson mass-shell condition  requires that
$S(p)$ be evaluated in a certain domain of complex $p^2$.  The required  
domain for meson decays and form factors can be
quite demanding.  To facilitate  such studies, we make use of an 
analytic parametrization of the numerical solutions of the quark DSE 
to represent $S(p)$ as an entire function in 
the complex $p^2$-plane describing absolutely confined~[\ref{RWK92}] 
dressed quarks.   Typically five parameters are used to achieve 
a good description of  pion and kaon observables:
$f_{\pi/K}$; $m_{\pi/K}$; $\langle\bar q q\rangle$; the $\pi$-$\pi$ 
scattering lengths; the charge radii $r_\pi$, $r_{K^0}$ and $r_{K^\pm}$;  
and the pion charge form factor~[\ref{R96},\ref{BRT96}].  Current efforts
along this line are concentrated upon vector and axial vector mesons
and hadronic and semi-leptonic decays.   Here we outline recent
studies of several electromagnetic transitions and also the $\rho\pi\pi$
coupling.

\medskip
\bcn
{\large\it 2.~The $\rho\pi\pi$ Coupling Constant}
\ecn

The first term in a skeleton graph expansion of the $\rho\pi\pi$ 
vertex~[\ref{T97},\ref{T96}] is
\beqar
\label{vint}
\lefteqn{\Lambda_\nu(P,Q)= -P_\nu  g_{\rho\pi\pi}\, F(Q^2) } \\
& & \nonumber
= 2 N_c\, {\rm tr}_s \int \frac{d^4k}{(2\pi)^4} 
 \Gamma_\pi(k^{\prime\prime\prime};-P_+) S(q^\prime)
            \Gamma^\rho_\nu(k^\prime;Q) S(q^{\prime\prime}) 
  \Gamma_\pi(k^{\prime\prime};P_-) S(q^{\prime\prime\prime})~,
\eeqar
where \mbox{$P_\pm = P \pm Q/2$} and  both pions are on 
the mass-shell.  The first argument of each BS amplitude is a relative
\mbox{$\bar q q$} momentum (we choose equal partitioning) and the second 
argument is the incoming meson momentum.   After the loop momentum
$k$ is specified in terms of one internal momentum, the others are 
easily deduced.  By definition we have \mbox{$ F(Q^2=-m_\rho^2)=1$}.  

For this $\rho\pi\pi$ study, we employ approximate 
$\Gamma_\pi$ and $\Gamma_\nu^\rho$ obtained from a rank-2 separable 
ansatz~[\ref{sep97}] for the ladder/rainbow kernel of the DSE and BSE.  
The $\pi$ is properly massless in the chiral limit as required by 
Goldstone's theorem; but the full consequences of the axial Ward-Takahashi 
identity~[\ref{MRT98}] are not preserved.   Parameters are fit to 
$m_{\pi/K}$ and $f_{\pi/K}$. The results have the form~[\ref{sep97}]
\beq
\Gamma_\pi (k;P) = i\gamma_5\; f(k^2)\, \lambda_1^\pi\, -
      \gamma_5 \,\Pslash \; f(k^2)\, \lambda_2^\pi~,
\label{sep_pi}
\eeq
\beq
\Gamma_\nu^\rho(k;P) = k_{\nu}^T \; g(k^2)\lambda_1^\rho 
+ i\gamma_{\nu}^T\; f(k^2)\lambda_2^\rho 
+ i\gamma_5 \epsilon_{\mu \nu \lambda \rho}\;
\gamma_{\mu} k_{\lambda} P_{\rho} \; g(k^2)\lambda_3^\rho~.
\label{vamp}
\eeq
Here \mbox{$g(k^2)=[A(k^2)-1]/a$} and \mbox{$f(k^2)=B(k^2)/b$} with 
$a$ and $b$ given in Ref.~[\ref{sep97}] from the quark DSE solution.
The relative strength of the $\lambda_i$ is given by the separable BSE 
solution which requires a phenomenological UV suppression.  In 
contrast to previous work~[\ref{sep97}], here we normalize the $\lambda_i$ 
in the canonical way~[\ref{MRT98}]
without such a suppression in any of the momentum dependent quantities.
\begin{table}[ht]
\caption{$g_{\rho\pi\pi}$ calculation and contributions from meson 
covariants. \label{tab:rpp} }
\vspace{0.2cm}
\begin{center}
\footnotesize
\begin{tabular}{|c|}\hline
 $g_{\rho\pi\pi}= 7.32$  [expt 6.05] \\
\end{tabular}\\
\begin{tabular}{|cl||cl|} \hline
  $\pi$ Covariants  &    & $\rho$ Covariants & \\ \hline
 $\gamma_5$ & 184\%  & $\gamma_\mu$ & 98.5\%   \\ 
 $\gamma_5 \gamma \cdot Q$ & -84\% & $\gamma_5 \epsilon_\mu \; 
                                           \gamma k Q$ & 1.4\% \\ 
            &      &     $k_\mu$   &     0.1\%  \\ \hline
\end{tabular}
\end{center}
\end{table}
The results for $g_{\rho\pi\pi}$ are given in \Table{tab:rpp} 
where an error in our previous work~[\ref{Tparis98}] is corrected.  
The empirical value associated with 
the \mbox{$\rho\rightarrow\pi\pi$} decay width is overestimated by 21\%.
The integral for $g_{\rho\pi\pi}$ is significantly influenced by the 
normalization of $\Gamma_\pi$; the form of the separable
amplitude \Eq{sep_pi} may be too primitive to generate a realistic norm.  
Improved BS amplitudes~[\ref{MR97},\ref{MT99a}] are available for 
future work.  The pseudovector $\pi$ component is much more important
here (-84\%) than for it is for $m_\pi$ 
and $f_\pi$ (25-35\%).  The sub-dominant $\rho$ amplitudes make only 
a minor correction.  

\medskip
\bcn
{\large\it 3.~The $\gamma^\ast \pi\rho$ Form Factor}
\ecn

The isoscalar $\gamma^\ast \pi\rho$ meson-exchange current contributes 
significantly to electron scattering from light nuclei. 
Our understanding of the deuteron EM structure functions
for \mbox{$Q^2 \approx 2-6~{\rm GeV}^2$} requires knowledge of this form 
factor~[\ref{VODG95}].   The general form of the vertex, and the explicit 
quark loop that arises in the impulse approximation generalized 
to include the dressing of the propagators and vertices, is
\beqar
\label{lam}
\lefteqn{ \Lambda_{\mu\nu}(P,Q)= -i\case{e}{m_{\rho}}\; \epsilon
_{\mu \nu \alpha \beta } \; P_{\alpha }Q_{\beta }\,g_{\gamma \pi\rho}\,
F(Q^2)  } \\
& & \nonumber
=\case{2N_c}{3}\,  {\rm tr}_s \int \frac{d^4k}{(2\pi)^4}
\, \bar{\Gamma}_\pi(k^{\prime\prime\prime};-P_+)\,
S(q^\prime)\,\Gamma_\nu(k^\prime;Q)\,
S(q^{\prime\prime})\, \Gamma_\mu^\rho(k^{\prime\prime};P_-)\, 
S(q^{\prime\prime\prime})~.
\eeqar
The momentum notation is the same as for the previous $\rho\pi\pi$ case. 
The dressed photon-quark vertex is taken to be the 
Ball-Chiu~[\ref{BC80}] ansatz 
\begin{equation}
\Gamma_\nu(k;Q)=-\case{i\gamma_\nu}{2} (A_++A_-)
+ \case{k_{\nu}}{k \cdot Q}\Bigl[ i\kslash
(A_- - A_+) + B_- -B_+ \Bigr]~, 
\label{ballchiu}
\end{equation}
where \mbox{$f_\pm = f(k_\pm)$} and \mbox{$k_\pm=k\pm \frac{Q}{2}$}.  
This form obeys the Ward-Takahashi 
identity (WTI) and the relevant symmetries and is conveniently 
determined completely in terms of the quark propagator.  It then follows
that \mbox{$Q_\nu\;\Lambda_{\mu\nu}=0$}; the $\gamma\pi\rho$ current is 
conserved.   With the above separable model $\pi$ and $\rho$ BS 
amplitudes, along with the associated quark propagator 
parameterization~[\ref{BRT96}], we obtain 
$g_{\gamma \pi \rho}=0.45$ in reasonable agreement with the
empirical value $g_{\gamma \pi \rho}^{\rm expt}=0.54\pm 0.03$ from $\rho$
decay. 
\begin{figure}[ht]
\ihsp \epsfig{figure=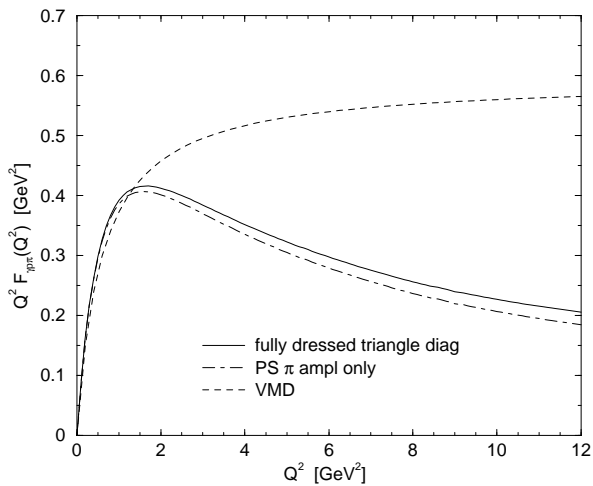,angle=0,height=6.5cm} \ihsp
\caption{The $\gamma^\ast \pi \rho $ transition form factor with
$\pi$ and $\rho$ on shell.
\label{fig:gprQ2FF} } 
\end{figure}

The pseudovector $\pi$ amplitudes $F_\pi$ and $G_\pi$ defined in 
\Eq{Gammapi} generate the correct asymptotic behavior of the pion charge 
form factor~[\ref{MR98}].  Also their UV relationship 
\mbox{$k^2 G_\pi(k^2)\rightarrow$} \mbox{$2F_\pi(k^2)$} implements 
convergence for the $f_\pi$ integral~[\ref{MR98}].  It is impossible 
to generate these properties in 
the previous separable ansatz approach; we use here the approximating 
forms~[\ref{MR98}] of a  numerical solution~[\ref{MR97}] 
\beq
E_\pi(k^2)\approx \frac{B_0(k^2)}{N_\pi}\, , \;\;  
F_\pi(k^2)\approx \frac{E_\pi(k^2)}{110 f_\pi}\, , \;\;  
G_\pi(k^2)\approx \frac{2 F_\pi(k^2)}{k^2 + M_{UV}^2}\, , \;\; 
H_\pi(k^2)\approx 0,
\label{MRpi-C7}
\eeq
where $N_\pi$ is fixed by the standard BS normalization 
condition~[\ref{MRT98}].  We use the quark propagator parameterization 
(set A)~[\ref{MR98}] associated with \Eq{MRpi-C7}.   
There is no dynamically matched $\Gamma_\mu^\rho$ available so we 
simply take  \Eq{vamp}.    We then obtain 
$g_{\gamma \pi \rho}=0.708$.  Since the canonical $\rho$ covariant is so
dominant, we expect that the mismatch is simply one of normalization and 
that the produced $F(Q^2)$ should be quite realistic. 
The result in \Fig{fig:gprQ2FF} is much softer than the
vector meson dominance (VDM) prediction. The available data for elastic 
EM deuteron form factors $A(Q^2)$ and $B(Q^2)$ in the range 
\mbox{$2-6~{\rm GeV}^2$} has been shown~[\ref{VODG95}] to strongly 
favor our previous $\gamma^\ast \pi\rho$ vertex result~[\ref{T96}]. 
The present work is less phenomenological, 
employs more realistic representations of the $\pi$ and $\rho$, and 
produces a harder form factor.  The influence on the deuteron form 
factors remains to be determined.
\begin{figure}[ht]
\ihsp \epsfig{figure=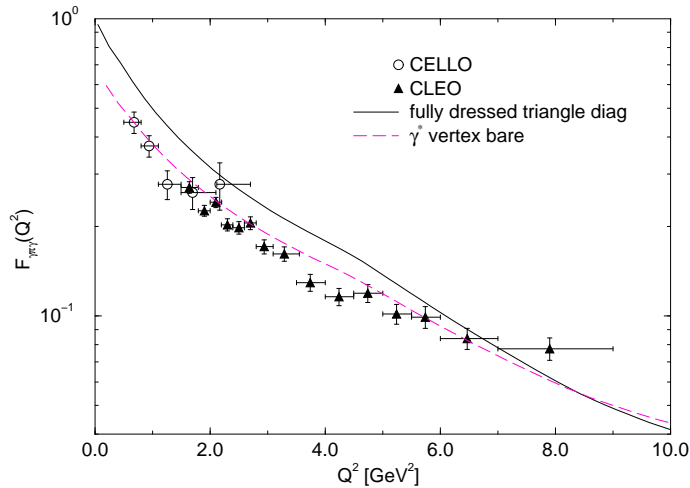,angle=0,height=6.5cm} \ihsp
\caption{The $\gamma^\ast \pi^0 \gamma $ transition form factor.  The
data are taken from Ref.~[\protect\ref{CELLO}] (CELLO) and 
Ref.~[\protect\ref{CLEO}] (CLEO). 
\label{fig:ggpFFlog} } 
\end{figure}

\medskip
\bcn
{\large\it 4.~ The $\gamma^\ast \pi^0 \rightarrow \gamma$ Transition}
\ecn

The coupling constant for the \mbox{$\pi^0 \rightarrow \gamma \gamma$} decay
is given by the axial anomaly and its value is a consequence of only 
gauge invariance and chiral symmetry in quantum field theory.  The form 
factor of this anomalous transition is not dictated by symmetries; 
it is of interest as a test of our ability to model nonperturbative QCD
because of the relatively simple hadronic dynamics that is involved.    
In the asymptotic UV region, one expects a simple result dictated by the 
known electromagnetic coupling to current quarks and the intrinsic 
properties of the pion.

The general form of the vertex allowed by CPT symmetry, and  the explicit 
quark loop that arises in the impulse approximation generalized 
to include the dressing of the propagators and vertices, are
\begin{eqnarray}
\label{int}
\lefteqn{ \Lambda_{\mu\nu}(P,Q)=i\case{\alpha }{\pi f_{\pi }}\,\epsilon
_{\mu \nu \alpha \beta }\,P_{\alpha }Q_{\beta }\, g_{\pi\gamma\gamma}\,
F(Q^2)  } \\
& & \nonumber
=\case{N_c}{3}\,  {\rm tr}_s \int \frac{d^4k}{(2\pi)^4}\,
S(q^\prime)\Gamma_\nu (k^\prime;Q) \,S(q^{\prime\prime})
\Gamma_\mu (k^{\prime\prime};-P-Q)\, S(q^{\prime\prime\prime})
\Gamma_\pi (k^{\prime\prime\prime};P) .
\end{eqnarray}
The $\gamma^\ast$ momentum is $Q$, and the other photon and the pion are on
the mass-shell.    With the convenient choice \mbox{$k^{\prime\prime\prime}=k$}, 
the other internal momenta can be determined.
The Ball-Chiu ansatz \Eq{ballchiu} is used for the dressed-quark-photon vertices 
$\Gamma_\nu$ and $\Gamma_\mu$.  The chiral limit anomalous  decay 
\mbox{$\pi^0 \rightarrow \gamma\gamma$} gives 
\mbox{$g^{\small 0}_{\pi\gamma\gamma}=1/2$} providing an excellent account 
of the $7.7~{\rm eV}$ width.     The form 
factor defined by \Eq{int} satisfies \mbox{$F(0)=1$}; it should not 
be confused with a different quantity
\mbox{$\tilde{F}(Q^2)=F(Q^2)/4\pi^2 f_\pi$} which contains the non-tensor 
strength of the transition matrix element 
\mbox{$M_{\mu \nu}= 2\Lambda_{\mu \nu}$} and in terms of which the 
CLEO~[\ref{CLEO}] data and some theoretical works are expressed.
\begin{figure}[ht]
\ihsp \epsfig{figure=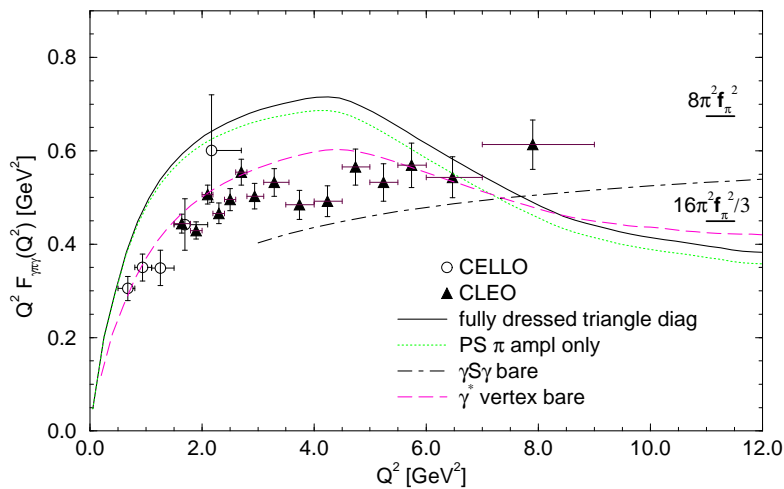,angle=0,height=6.5cm} \ihsp
\caption{The $\gamma^\ast \pi^0 \gamma $ transition form factor 
times $Q^2$.  The asymptotic limits marked are described in the text.  
\label{fig:ggpQ2FF} } 
\end{figure}

Here we update an earlier study~[\ref{FMRT95}] by using
the more realistic $\Gamma_\pi$ in \Eq{MRpi-C7} along with the associated 
quark propagator (set A)~[\ref{MR98}].   We obtain 
\mbox{$g_{\pi\gamma\gamma}=0.4996$} at the physical $m_\pi$ value.  
The form factor is displayed in \Fig{fig:ggpFFlog}. 
Although the anomalous coupling strength is correct, the 
calculation does not fall off fast enough  in the infrared and
significantly overestimates the data.   It is unlikely that our description
of the pion is responsible for this; the sub-dominant 
amplitudes $F_\pi$, $G_\pi$ give a negligible contribution here
while the closely related dynamical quantities, $r_\pi$ and $f_\pi$, are
well described~[\ref{MR98}].  The removal of dressing at the
virtual photon vertex produces the dashed line.  The better agreement 
with the data is rather fortuitous since the coupling constant is 
reduced to 70\% of the former correct value.  The pion transition radius 
from the impulse approximation ($\sim 0.48~{\rm fm}$) is clearly less than
that suggested by the data ($\sim 0.65~{\rm fm}$);  a similar underestimate
also occurs for the pion charge radius in this approach.

Our results for $Q^2 F(Q^2)$ are displayed in \Fig{fig:ggpQ2FF}.  
The approach to the asymptotic limit of the present dressed quark 
loop \Eq{int} is quite slow and is governed by the following 
considerations.  After the mass-shell conditions for 
the pion and the real photon are realized, one finds that in the domain 
\mbox{$k^2 < 1~{\rm GeV}^2 \ll Q^2$} of integral support dictated by 
$\Gamma_\pi(k^2)$,  the leading behavior of the momenta for each  quark
propagator is: 
\mbox{$(q^\prime)^2,~(q^{\prime\prime\prime})^2 \sim {\cal O}(kQ)$}, and 
\mbox{$(q^{\prime\prime})^2 = \frac{Q^2}{2}[1 + {\cal O}(k/Q)]$}.  Using 
\mbox{$A(p^2)=1+{\cal O}(1/p^2)$} and \mbox{$B(p^2)\sim{\cal O}(1/p^2)$}, 
one finds from \Eq{ballchiu} that the leading behavior of each of the 
photon vertices  is \mbox{$\Gamma_\nu =$}
\mbox{$-i\gamma_\nu [1+ {\cal O}(1/kQ)]$}.    The photon insertions 
of the diagram effectively collapse to a point axial vector with the
\mbox{$\Gamma_\nu S(q^{\prime\prime})\Gamma_\mu$} leg of the loop having
the leading form 
\beq
\frac{2i}{Q^2}\; \gamma_\nu \qslash^{\prime\prime} \gamma_\mu = 
\frac{2i}{Q^2} \; \epsilon_{\mu \nu \alpha \beta} \; \gamma_5 
\gamma_\alpha Q_\beta + \cdots~,
\label{gSg}
\eeq
where the current mass is ignored and the terms not shown do not survive 
the spin trace.   The loop integral coupling the $\pi$ to 
$\gamma_5 \gamma_\alpha$ gives exactly $f_\pi P_\alpha/3$
by definition~[\ref{MRT98}].  This gives the asymptotic limit 
\mbox{$\Lambda_{\mu \nu} (Q^2)\rightarrow$}
\mbox{$ \frac{2f_\pi}{Q^2} \; i\epsilon_{\mu \nu \alpha \beta}\; P_\alpha Q_\beta [1 + {\cal O}(1/Q)]$} or the form factor limit 
\mbox{$\frac{16}{3}\pi^2f_\pi^2 + {\cal O}(1/Q)$}, in agreement 
with Refs.~[\ref{KK99},\ref{CDR99}].  The limit marked by 
\mbox{$ 8 \pi^2 f_\pi^2$} is from pQCD factorization~[\ref{LB80}]. 
It is evident that the sub-dominant amplitudes $F_\pi$ and $G_\pi$ make 
a minor contribution although the integrated effect via
the produced $f_\pi$ value is some 30\%. 

With the hard leg \mbox{$\Gamma_\nu S(q^{\prime\prime})\Gamma_\mu$} 
taken to
be bare the result is the dot-dashed line in \Fig{fig:ggpQ2FF}.  This is 
consistent with the evident sub-leading correction in the propagator denominator 
being \mbox{$\langle k/Q \rangle \sim 30{\rm \%}$} at 
\mbox{$Q^2 \sim 10~{\rm GeV}^2$}.    The dressing of the hard propagator 
$S(q^{\prime\prime})$ contributes little to the dot-dashed curve; thus the 
difference 
between the three relevant curves illustrates the persistent contribution 
from photon vertex dressing.    Since the Ball-Chiu vertex is exact at 
both \mbox{$Q^2=0$} and the UV limit, and only the longitudinal component 
is correct for all $Q^2$, it is possibly the deficiencies of this ansatz 
at infrared and intermediate momenta that are being exposed in the 
present study.   This is also the preliminary finding from a study of the 
ladder Bethe-Salpeter solution for the vector vertex~[\ref{MT99b}].  
 
\medskip
\bcn
{\large\it 5.~Summary}
\ecn
 
The results presented here 
suggest that the present approach to modeling low-energy QCD  can capture 
the mechanisms that dominate infrared physics.  The sub-dominant pseudovector 
terms in the pion BS amplitude make a much stronger contribution to
$g_{\rho\pi\pi}$ than they do to $m_\pi$, $f_\pi$;  the contributions to 
the $\gamma^\ast \pi^0 \gamma $ and  $\gamma^\ast \pi \rho$ transition 
form factors are minor at low and intermediate momenta.  The new result here
for the $\gamma^\ast \pi \rho$ form factor should be applied to the deuteron
electromagnetic form factors where this meson exchange current is still a
serious ambiguity.  Our examination of the pion axial anomaly form factor 
with the  Ball-Chiu ansatz for the dressed photon-quark vertex clarifies 
the slow approach to asymptotic behavior and suggests that deficiencies 
in this ansatz at low momenta may be evident.  A study of electromagnetic 
radii should be made from this perspective. 

\smallskip
\bcn
Acknowledgments
\ecn
Thanks are due to the conference organizers, especially Dubravko Klabucar, 
for the excellent hospitality and enjoyable atmosphere provided at this 
conference.   Helpful discussions with D. Klabucar, C. D. Roberts, 
P. Maris and M. Pichowsky are gratefuly acknowledged.   This work is 
partly supported by the US National Science Foundation under grants 
No. PHY97-22429 and INT96-03385.


\bcn
References
\ecn
\begin{enumerate}

\item \label{DSErev} C. D. Roberts and A. G. Williams, Prog. Part. Nucl. 
Phys.{\bf 33} (1994) 477. \vspace*{-0.5\baselineskip}

\item \label{T97} P. C. Tandy, Prog. Part. Nucl. Phys. {\bf 39} (1997) 
117. \vspace*{-0.5\baselineskip} 

\item \label{MR97} P. Maris and C. D. Roberts, Phys. Rev. {\bf C56} (1997)
3369. \vspace*{-0.5\baselineskip}

\item \label{MRT98} P. Maris, C. D. Roberts and P. C. Tandy, Phys. Lett. 
{\bf B420} (1998) 267. \vspace*{-0.5\baselineskip}

\item \label{RWK92} C. D. Roberts, A. G. Williams and G. Krein, Int. J. 
Mod. Phys. {\bf A7} (1992) 5607. \vspace*{-0.5\baselineskip}

\item \label{R96} C. D. Roberts, Nucl. Phys. {\bf A605} (1996) 
475. \vspace*{-0.5\baselineskip}

\item \label{BRT96} C. J. Burden, C. D. Roberts and M. J. Thomson, Phys. 
Lett. {\bf B371} (1996) 163. \vspace*{-0.5\baselineskip}

\item \label{T96} P. C. Tandy, Prog. Part. Nucl. Phys. {\bf 36} (1996) 97;
 K. L. Mitchell, PhD Thesis, Kent State University, unpublished, 
(1995). \vspace*{-0.5\baselineskip}

\item \label{sep97} C. J. Burden, Lu Qian, C. D. Roberts, P. C. Tandy and 
M. J. Thomson, Phys. Rev. {\bf C55} (1997) 2649. 
\vspace*{-0.5\baselineskip}

\item \label{Tparis98} P. C. Tandy, ``Modeling nonperturbative QCD for 
mesons
and couplings'', Proceedings, IV$^{th}$ Workshop on Quantum Chromodynamics,
June, 1998, nucl-th/9812005, to be published. \vspace*{-0.5\baselineskip}

\item \label{MT99a} P. Maris and P. C. Tandy, "Vector meson masses and 
decay constants", preprint KSUCNR-104-99, in preparation, 
(1999). \vspace*{-0.5\baselineskip}

\item \label{VODG95} J. W. Van Orden, N. Devine and F. Gross, Phys. Rev. 
Lett. {\bf 75} (1995) 4369. \vspace*{-0.5\baselineskip}

\item \label{BC80} J. S. Ball and T. W. Chiu, Phys. Rev. {\bf D22} (1980)
2542. \vspace*{-0.5\baselineskip}

\item \label{MR98} P. Maris and C. D. Roberts, Phys. Rev. {\bf C58} (1998)
3659. \vspace*{-0.5\baselineskip}
 
\item \label{FMRT95} M. R. Frank, K. L. Mitchell, C. D. Roberts and P. C. 
Tandy, Phys. Lett. {\bf B359} (1995) 17. \vspace*{-0.5\baselineskip}

\item \label{CLEO} J. Gronberg \etal (CLEO Collaboration), Phys. Rev. 
{\bf D57} (1998) 33. \vspace*{-0.5\baselineskip}

\item \label{CELLO} H.-J. Behrend \etal (CELLO Collaboration), Z. Phys. 
{\bf C49} (1991) 401. \vspace*{-0.5\baselineskip}

\item \label{LB80} G. P. Lepage and S. J. Brodsky, Phys. Rev. {\bf D22} 
(1980) 2157.  \vspace*{-0.5\baselineskip}

\item \label{KK99} D.~Kekez and D.~Klabucar, ``$\gamma^\ast$ $\gamma$
$\to$ $\pi^0$ transition and asymptotics of $\gamma^\ast \gamma$ and
$\gamma^\ast \gamma^\ast$ transitions of other unflavored pseudoscalar
mesons," hep-ph/9812495; and these proceedings. \vspace*{-0.5\baselineskip}

\item \label{CDR99} C.~D.~Roberts, ``Dyson Schwinger equations: 
connecting small and large length-scales,'' hep-ph/9901091, these 
proceedings. \vspace*{-0.5\baselineskip}

\item \label{MT99b} P.~Maris, private communication, 
(1999). \vspace*{-0.5\baselineskip}

\end{enumerate}

\end{document}